\documentclass[referee]{raa}
\usepackage{graphicx,times}
\usepackage{natbib}
\usepackage{amssymb,amsmath}
\bibpunct{(}{)}{;}{a}{}{,}

\usepackage[a4paper=true,dvipdfm=true,pagebackref=true]{hyperref}
\hypersetup{pdftitle = The title of my PDF, pdfauthor = My name, pdfsubject= The subject, pdfkeywords = keyword1 keyword2 keyword3}
\hypersetup{colorlinks = true, linkcolor = green, anchorcolor = red, citecolor = blue, filecolor = red, pagecolor = red, urlcolor = red}

\begin{document}

   \title{Effects of Self-gravity on Mass-loss of the Post-impact Super-Earths}

   \author{Jiang Huang\inst{1,2}, Wei Zhong\inst{1,2}, Cong Yu\inst{1,2}}

   \institute{ School of Physics and Astronomy, Sun Yat-Sen University, Zhuhai 519082,  People's Republic of China {\it yucong@mail.sysu.edu.cn}\\
   \and
   CSST Science Center for the Guangdong-Hong Kong-Macau Greater Bay Area, Zhuhai 519082, People's Republic of China}

\abstract{Kepler's observations show most of the exoplanets are super-Earths. The formation of super-Earth is generally related to the atmospheric mass loss that is crucial in the planetary structure and evolution.  The shock driven by the giant impact will heat the planet, resulting in the atmosphere escape.  We focus on whether self-gravity changes the efficiency of mass loss. Without self-gravity, if the impactor mass is comparable to the envelope mass, there is a significant mass-loss. The radiative-convective boundary will shift inward by self-gravity. As the temperature and envelope mass increase, the situation becomes more prominent, resulting in a heavier envelope.  Therefore, the impactor mass will increase to motivate the significant mass loss, as the self-gravity is included. With the increase of envelope mass, the self-gravity is particularly important.
}

   \authorrunning{Huang et al. }            
   \titlerunning{Effects of self-gravity on giant impact}  
   \maketitle

%
\section{Introduction} \label{sect:intro}

The observation of the Kepler reveals a large number of short-period planets with size between Earth and Neptune, referred to as the super-Earth and mini-Neptune \citep{2010yCatp021033001H,2011yCat..17360019B,2013ApJ...766...81F,2013PNAS..11019273P,2013Sci...340..587B,2016ApJ...822...86M}. The atmospheric mass of these super-Earth is about 10 \% of that of the core mass \citep{2008ApJ...673.1160A,2014ApJ...792....1L}, implying the evolution is undergoing a significant mass loss.  Therefore, some feasible mechanisms are required to explain the formation of gas giant or super-Earth.

Super-Earths may be formed by migrating inwardly for the gravitational interaction of the protoplanetary disk \citep{2007ApJ...654.1110T,2008MNRAS.384..663R,2010ApJ...724L..53S,2010MNRAS.401.1691M}, or super-Earths are formed in situ by various mechanisms. We focus on the in situ formation of super-Earth. The super-Earth may be formed by the entropy advection \citep{2020MNRAS.494.2440A}, pebble accretion and isolation  \citep{2019A&A...632A...7L,2020ApJ...896..135C,2020RAA....20..164L} and tidally forced turbulence \citep{2017ApJ...850..198Y}. In the dispersing disk, the planet's core may be in the magma ocean stage \citep{2019ApJ...887L..33K}. The ultra-hot core heats the envelope and then forces the atmosphere to escape \citep{2021tsc2.confE..50M}. If the core energy is large enough, a bare core can be reminded. However, the entire envelope cannot be stripped without enough energy. Thus, a gas-rich sub-Neptune will be formed \citep{2016ApJ...825...29G}. In addition, the atmosphere of the close-in planet will be strongly affected by photoevaporation \citep{2013ApJ...775..105O,2017ApJ...847...29O} and the magnetorotational instability (MRI, \cite{2013ApJ...769...76B}). A Parker wind will blow off the atmosphere. 

The photoevaporation and core-powered effect are sensitive in a dispersing disk instead of a gas-rich phase \citep{2013ApJ...776....2L,2014ApJ...795...65J,2017AJ....154..109F}. However, the evolutionary timescale of the photoevaporates is much long for several Gyr. Thus, the effect of a giant impact is even more significant \citep{2015MNRAS.448.1751I}. In the final stages of planetary formation, the giant impact can significantly or even completely remove the atmospheric envelope \citep{2015ApJ...812..164L,2015Icar..247...81S,2016ApJ...817L..13I}. A shock will be induced by the giant impact \citep{2015Icar..247...81S,2015MNRAS.448.1751I,2016ApJ...825...29G,2019ESS.....430005B}, which will heat the planet resulting in a significant mass loss. However, the thermal structure of \cite{2019ESS.....430005B} neglected the effect of the self-gravity.

In the core-accretion frame, self-gravity can be ignored in the initial stages of planetary accretion \citep{2019MNRAS.487.2319B}. But as planets enter a runway accretion process, the effects of self-gravity cannot be ignored \citep{2019MNRAS.490.3144B}. The close-in planets can form the low-frequency nonradial oscillations by the dynamic tide, more precisely called the gravito-inertial waves (i.e., g-mode). Taking the tides into account, the competition between the self-gravity and tides will introduce tidal disruption \citep{2021A&A...652A.154D}. Planetary rotation produces r-mode. Considering self-gravity, the interplay between the two waves would alter the planet's shape \citep{2019MNRAS.488.1960L}. Self-gravity, therefore, is crucial in the formation of planets.

We investigate the effect of self-gravity on the giant impact. During impact, the kinetic energy of the impactor is converted to heat the planet. Thus, the envelope is heated and inflated, resulting in mass loss.  Self-gravity increases with the atmospheric mass. The thermal structure may be changed by the self-gravity, in which the radiative-convective boundaries (RCBs) may be shifted inward \citep{2017ApJ...850..198Y}. Compared with the case of ignoring self-gravity, the corresponding impactor mass would be increased to support significant mass loss.

This work is listed as follows: Section \ref{sect:structure and evolution} shows that a model of the planet after a major impact was constructed and evolved with consideration of envelope's self-gravity. In Section \ref{sect:results}, we calculate the atmospheric loss results under different parameters. The conclusions and discussion are listed in Section \ref{sect:conclusions}.

\section{The planetary structure after the giant impact}\label{sect:structure and evolution}

In general, giant impact processes involve the loss of atmospheric mass on planets. In addition, the post-impact planetary structure depends on the thermal state of the envelope. In this work, we investigate how self-gravity changes the thermal structure. We construct a post-impact model with self-gravity for the time-dependent evolution of the planet. In section \ref{structure}-\ref{energy}, we list the structure functions, outer boundaries and energy as follows.

\subsection{H/He Envelope Structure Model}\label{structure}

We construct a two-layer model with an interior adiabatic convective zone and an exterior radiative zone \citep{2006ApJ...648..666R,2014ApJ...786...21P,2015MNRAS.448.1751I}. The atmospheric envelope is
roughly homogeneous around the core, which is approximately spherically symmetric for calculation. The H/He envelope structure is governed by the following equations of mass conservation, hydrostatic equilibrium, and the temperature gradient  \citep{2012sse..book.....K,2017ApJ...850..198Y}:
\begin{equation}
\frac{dM_{\rm r}}{dr}=4\pi r^{2}\rho,
\label{mass_conservation}
\end{equation}
\begin{equation}
\frac{dP}{dr}=- \frac{GM_{\rm r}}{r^{2}}\rho.
\label{hydrostatic_equilibrium}
\end{equation}
\begin{equation}
\frac{dT}{dr}= \frac{T}{P}\frac{dP}{dr}\nabla,
\label{temperature_gradient}
\end{equation}
where $M_{\rm r}$ is the total mass with core mass $M_{\rm c}$ and the envelope mass $M_{\rm env}$. $r$, $\rho$, $P$, $T$ and $\nabla$ represent the radius, density, pressure, temperature, and temperature gradient, respectively. Meanwhile, the gas satisfies the ideal gas law, $P=R\rho T/\mu$ with molar gas constant $R$, the mean molecular weight $\mu $. We set $\mu =2.3$. Through the above equation, the convective profile can be determined. In particular, temperature gradient depends on the radiative and adiabatic gradient, i.e., $\nabla = \min(\nabla_{\rm rad},\nabla_{\rm ad})$. The structure is divided into the radiative and convective parts.

\subsubsection{The radiative structure}\label{radiative_structure}

The radiative layer is approximately treated as an isothermal zone, in which the temperature seems equal to the outer boundary. To simplify the simulation, we neglect the radiative gradient. The temperature at the radiative-convective boundary (RCB) relates to the outer temperature, i.e., which is approximately the equilibrium temperature \citep{2019ESS.....430005B}:
\begin{equation}
T_{\rm eq}=[\frac{1-A_B}{4}]^{\frac{1}{4}}\sqrt{\frac{R_{\rm star}}{a}}T_{star},
\label{t_eq}
\end{equation}
where $A_B$ is the Bond albedo, $R_{\rm star}$ is the stellar radius, and $T_{star}$ is the star's effective temperature. The radius and temperature of the host star are $R_{\rm star}=R_{\odot }$, $T_{\rm star}=5.77\times 10^{3} K$, respectively. In this layer, we choose $r$ as the independent variable, and the equation (\ref{temperature_gradient}) will be removed. We will solve the equations (\ref{mass_conservation})-(\ref{hydrostatic_equilibrium}) with $r=R_{\rm out}$.

\subsubsection{The adiabatic structure}\label{adibatic_structure}

The temperature at RCBs is determined by the outer temperature since the radiative temperature is $T_{\rm RCB}=T_{\rm eq}$. We set temperature as the argument, thus, the equations (\ref{mass_conservation})-(\ref{temperature_gradient}) will satisfy
\begin{equation}
\frac{dr}{dT}=\frac{1}{GM_{\rm r}},
\label{convective_radius}
\end{equation}
\begin{equation}
\frac{dM_{\rm r}}{dT}=-\frac{4\pi r^{4}}{GM_{\rm r}}\frac{P}{T}
\label{convective_mass}
\end{equation}
\begin{equation}
\frac{dP}{dT}= \frac{P}{T}\frac{1}{\nabla_{\rm ad}},
\label{convective_pressure}
\end{equation}
where $\triangledown_{ad}$ is determined by the adiabatic index, and $\gamma$ is the adiabatic index of the gas \citep{2006ApJ...648..666R}. For monatomic gas $\nabla_{\rm ad}=2/5$, and for diatomic, e.g., $H_{2}$, $\nabla_{\rm ad}=2/5$, which is $\gamma=7/5$.

\subsection{The outer boundary conditions}

The outer boundary is determined by the lesser of the Bondi radius and the Hill radius \citep{2017ApJ...850..198Y}:
\begin{equation}\label{equation:1}
R_B\approx 90R_{\bigoplus } \left [ \frac{(1+GCR)M_c}{5M_{\oplus }} \right ]\left ( \frac{1000K}{T} \right ),
\end{equation}
\begin{equation}\label{equation:2}
R_H\approx 40R_{\bigoplus } \left [ \frac{(1+GCR)M_c}{5M_{\bigoplus }} \right ]^{1/3}\left ( \frac{a}{0.1 au} \right ),
\end{equation}
where $M_{\rm c}$ is the mass of the core, $a$ is orbital semi-major axis of planet, $GCR=M_{\rm env}/M_c$ is the mass ratio of envelope to core.

\subsection{The energy for the giant impact}\label{energy}

The total energy including the thermal and potential energy determines the planetary evolution. The planet model we construct is a young planet with a significant H/He envelope. Under such conditions, the planet still retains much thermal energy. Thus, the base temperature of the planetary envelope is higher than the melting point of silicon. Assuming that the core is in a completely molten state, and the heat conduction is effective between the core and the base of the envelope, so that the core temperature and the base temperature is approximate, $T_c\approx T_b$.

 For the terrestrial planets, the internal adiabatic core is at a depth of several thousand kilometers, and the temperature change is slight \citep{2010PEPI..183..212K}. Therefore, we consider an isothermal core, and estimate core energy:
\begin{equation}\label{equation:13}
E_c\sim c_{\rm v,c} T_{\rm c} M_{\rm c},
\end{equation}
where $c_{\rm v,c}\sim5-10\times 10^6 erg\,g^{-1} K^{-1}$ describes the specific heat capacity of the planet's core \citep{2001PhRvB..64d5123A,2012ApJ...761...59L}. We employ $k_B/[\mu_{\rm c} (\gamma_{\rm c}-1)]$  to estimate the value of $c_{\rm v,c}$. Note that $\mu_c$ and $\gamma_c$ are the mean molecular weight and adiabatic index of the core, respectively. In this work, we set $c_{\rm v,c}=7.5\times10^6 erg\,g^{-1} K^{-1}$.

The envelope mass is mainly distributed in convective zone, and the mass of radiative zone is negligible. Therefore, we mainly discuss the gravitational potential and thermal energy of the convective zone, in which can be shown as follows:
\begin{equation}
E_{\rm gra}=4\pi G\int_{R_{\rm c}}^{R_{\rm rcb}}M_r\rho r^{2}dr,
\label{gravity}
\end{equation}
and
\begin{equation}
E_{\rm th}=4\pi \frac{k_{\rm B}}{\mu (\gamma -1)}\int_{R_{\rm c}}^{R_{\rm rcb}}\rho T r^{3}dr.
\label{thermal_energy}
\end{equation}

After a giant impact, the envelope will be heated and inflated, resulting in significant mass-loos. As it gradually cools and shrinks, the outer density and mass loss rate will be decreased. The mass-loss rate is approximately
 \citep{2016ApJ...817..107O}:
\begin{equation}\label{equation:16}
\dot{M}_{\rm env} = -4\pi R_{\rm out}^2 \rho_{\rm out} c_{\rm s},
\end{equation}
where $c_s=\sqrt{\gamma_{\rm disk} k_{\rm B} T_{\rm eq}/\mu }$ is the sound speed with $\gamma_{\rm disk} = 1.0$. 
 The mass-loss rate, driven by the Parker wind, was characterized by the density at the boundary. The mass of the envelope is concentrated in RCB. The energy with a given mass-loss rate is required
\begin{equation}\label{equation:17}
\dot{E}_{\rm env,m} \approx \frac{GM_{rcb}\dot{M}_{env}}{R_{rcb}}.
\end{equation}
The planetary evolution is a cooling process. The corresponding luminosity at $R_{\rm rcb}$ can be determined by the radiative gradient, and can be written as follows:
\begin{equation}\label{equation:18}
\dot{E}_{\rm env,L} = -L_{\rm rcb}=-\nabla_{\rm ad}\frac{64\pi \sigma T_{\rm rcb}^3GM_{\rm rcb}\mu }{3\kappa _R\rho _{\rm rcb}k_B},
\end{equation}
where $\sigma$ is the Stefan-Boltzmann constant, $\kappa _R$ is the Rosseland mean opacity of the envelope at $R_{rcb}$. We set $\kappa _R=0.1cm^2g^{-1}$ \citep{2008ApJS..174..504F}. When the mass is lost from the outer edge of an envelope, the energy is required to support the same thermal profile. The luminosity at the RCB $L_{\rm rcb}$ gives the upper limit of the evolution, so that the maximum mass-loss rate can be determined by
\begin{equation}\label{equation:19}
\dot{M}_{\rm env,max} \approx  -\frac{L_{\rm rcb}R_{\rm rcb}}{GM_{\rm c}}.
\end{equation}
We ignore the energy losses of radiative process, and mass-loss is derived by the cooling luminosity. Thus, the mass-loss rate can be regarded as the absolute upper limit.

\section{RESULTS}\label{sect:results}

 We have constructed a planetary model after a giant impact and explored their evolution. The giant impact will heat the core and atmospheric envelope, resulting in a significant mass loss. We have shown the planetary structure with the self-gravity in Section \ref{strucure}, and the effect of the self-gravity on the mass-loss in the Section \ref{self-gravity}

 \subsection{The structure after the giant impact}\label{strucure}

\begin{figure*}
\centering
\includegraphics[width=7.4cm]{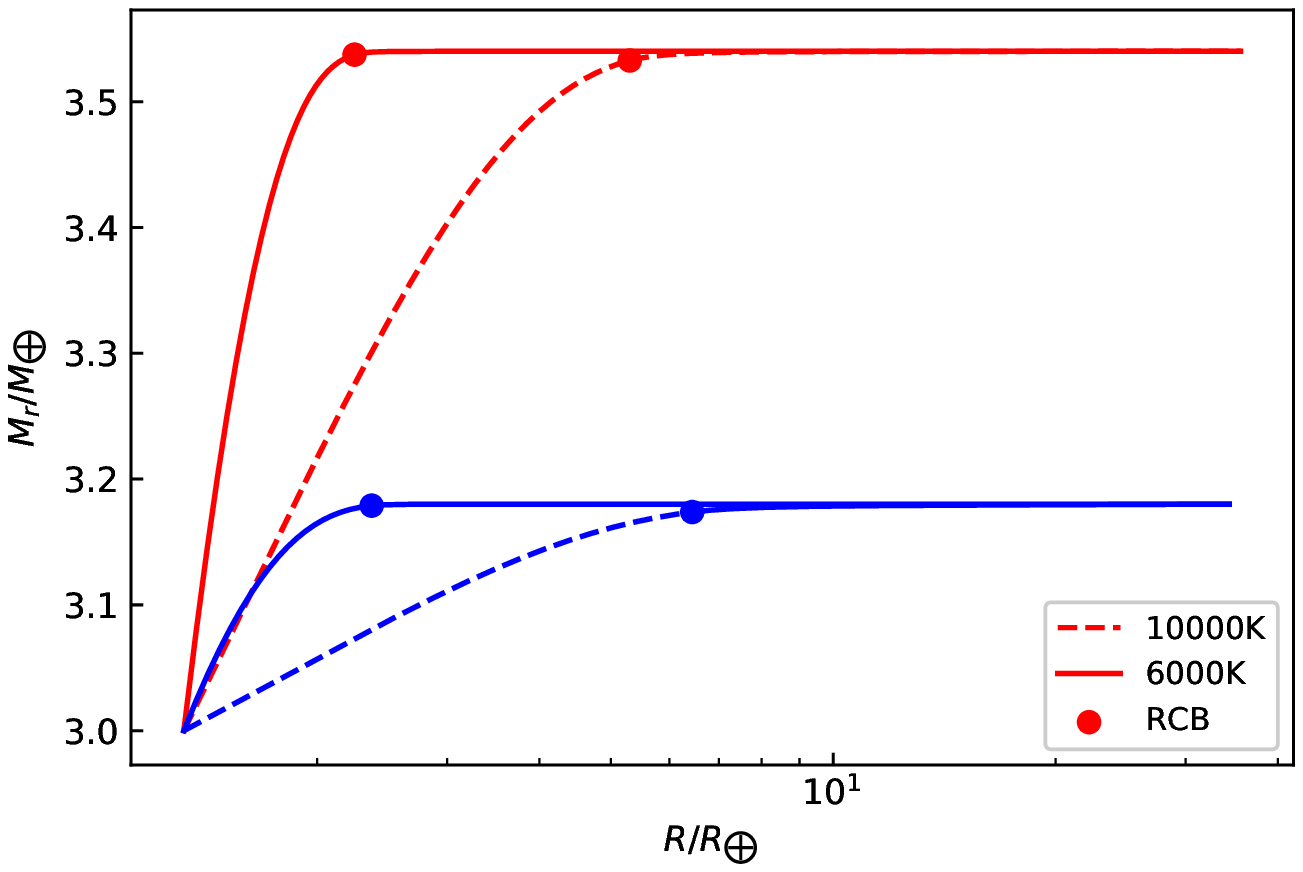}
\includegraphics[width=7.4cm]{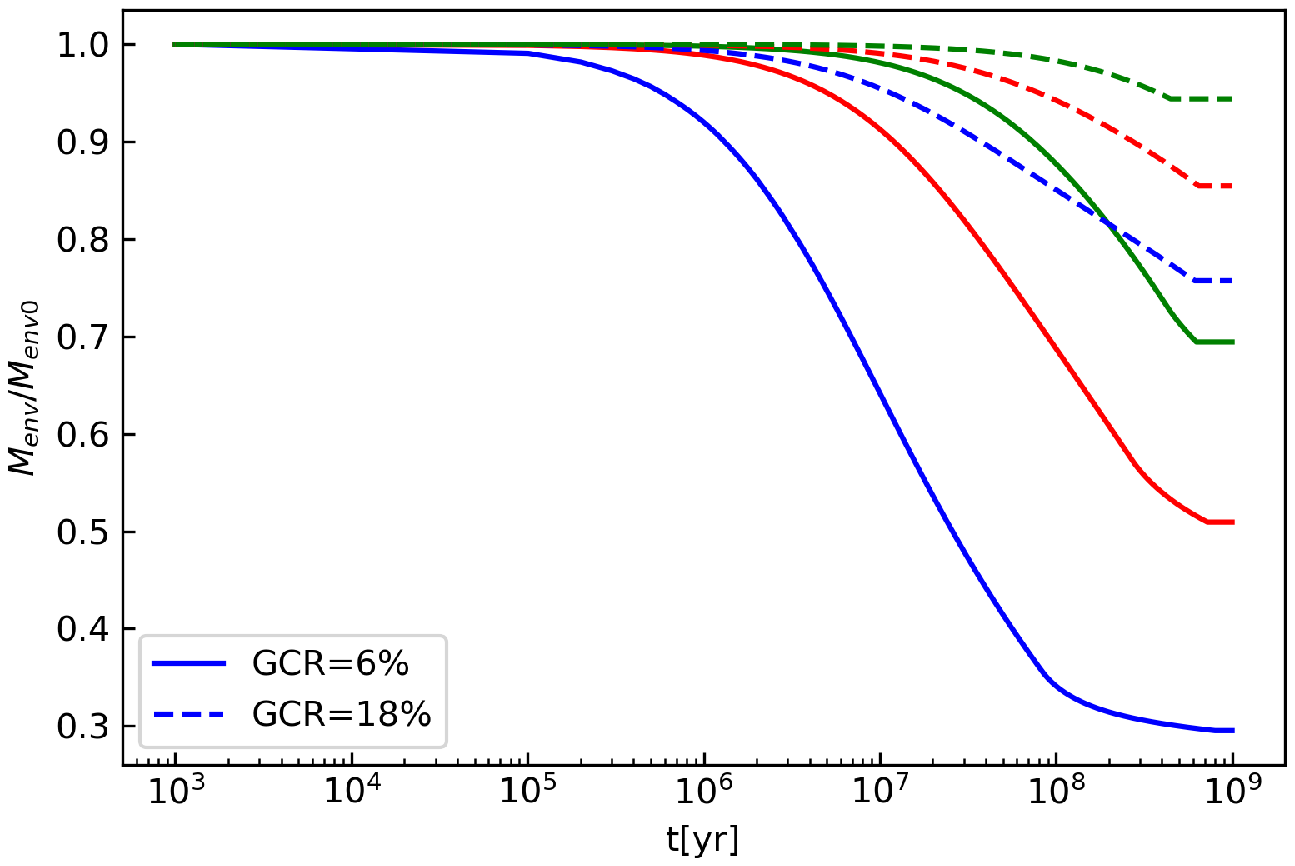}
\includegraphics[width=7.4cm]{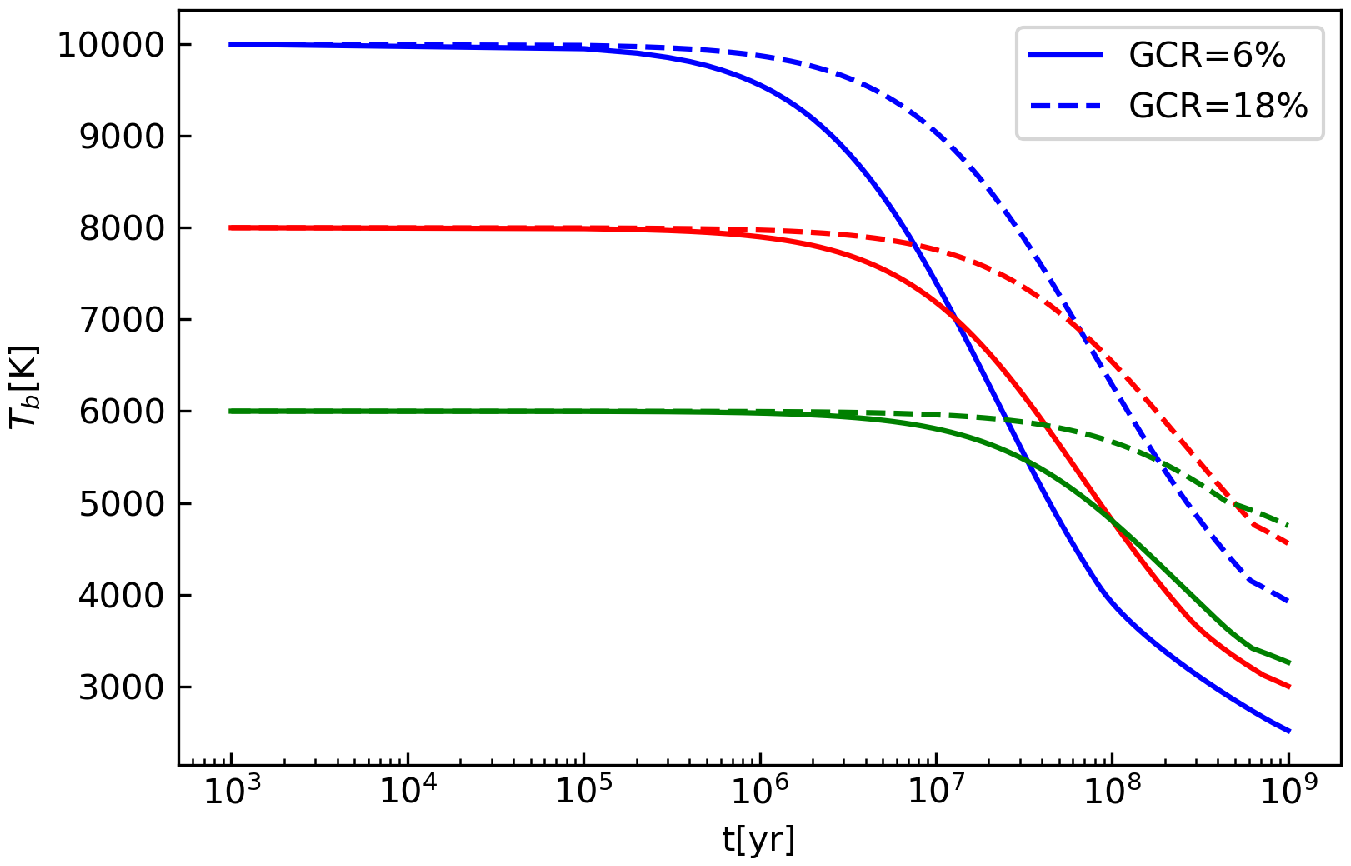}
\includegraphics[width=7.4cm]{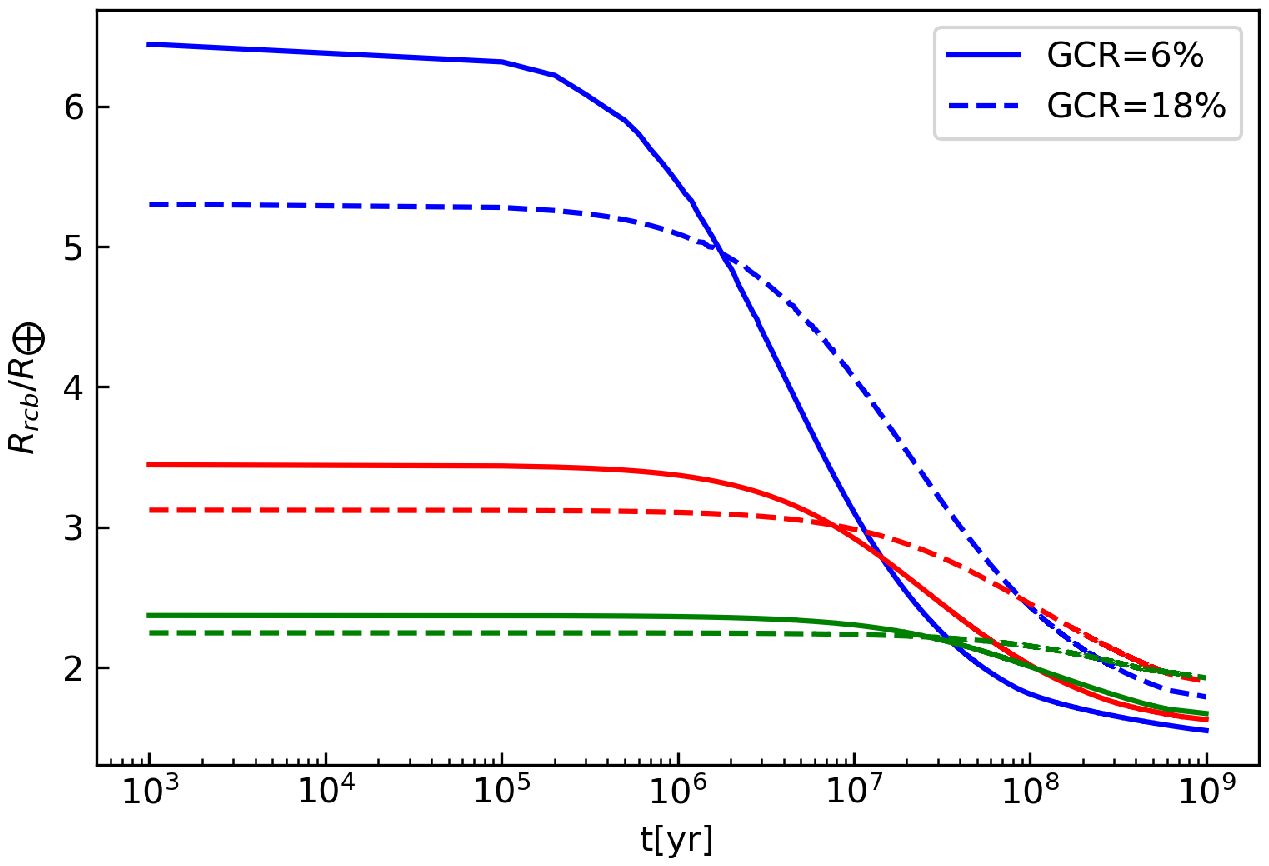}	
		
\caption{ The upper left:radial mass profile. Solid and dashed lines represent the cases of $T_{\rm b}=6000K$ and $10,000 K$, respectively. The blue and red lines correspond to the case of $GCR=6\%$ and 18\%, respectively. The upper right: the evolution of the atmospheric mass. The lower panels: the evolutions of temperature and radius at the RCBs from left to right. Solid and dashed lines represent the cases of $GCR_{\rm 0}=6\%$ and $18\%$, respectively. Green, red and blue lines corresponds to the initial temperatures of $T_{\rm b,0}=$ 6000, 8000, and 10,000 K at the base of the envelope, respectively.
}
\label{Fig:strucutre_evolution}
\end{figure*}

Following Sections \ref{adibatic_structure} and \ref{radiative_structure}, we can obtain the radial profile of the envelope shown in Figure \ref{Fig:strucutre_evolution}. We employ the odeint function provided by Scipy to integrate the structural equations from the outer boundary to the interior layer. The core mass and radius are set to $M_c=3M_{\oplus }$ and $R_c=1.32R_{\oplus }$, respectively. The planet is located at $0.1 au$.

The density $\rho_{\rm b}$, pressure $P_{\rm b}$, and temperature $T_{\rm b}$ at the base of the envelope would change the planetary structure. As shown in the upper left of Figure \ref{Fig:strucutre_evolution}, the envelope masses at the RCBs increase with the total mass (gas-to-core ratio, GCR). Besides, the RCB also shifts inward because the self-gravity increases with the envelope mass. In addition, the RCB will be pushed outward when the base temperature increases, which means the evolutionary timescale will be reduced, and the internal pressure pushes the materials to escape the Hill radius. Therefore, a significant mass loss will occur in the higher base temperature.

According to Section \ref{energy}, the evolution of mass, temperature and the radius at RCBs are also shown in Figure \ref{Fig:strucutre_evolution}. The mass-loss rate refers to the initial gas-to-core ratio $GCR_{\rm 0}$ and temperature $T_{\rm b,0}$ at the base of the envelope. Under the same $T_{\rm b,0}$, self-gravity increases with the initial GCR. Thus, the mass-loss rate is reduced (the upper right), and then, the evolutionary profiles of the temperature will be flatten. However, the location of the RCBs was pushed outward (the lower right). In addition, when a planet with the same initial GCR, the efficiency of the mass-loss would increase with the $T_{\rm b,0}$. In particular, the tendency of the decrease in the base temperature flattens (the lower left). The RCBs will be pushed inward.

\subsection{The effect of the self-gravity}\label{self-gravity}
\begin{figure}
   \centering

  \includegraphics[width=7.42cm]{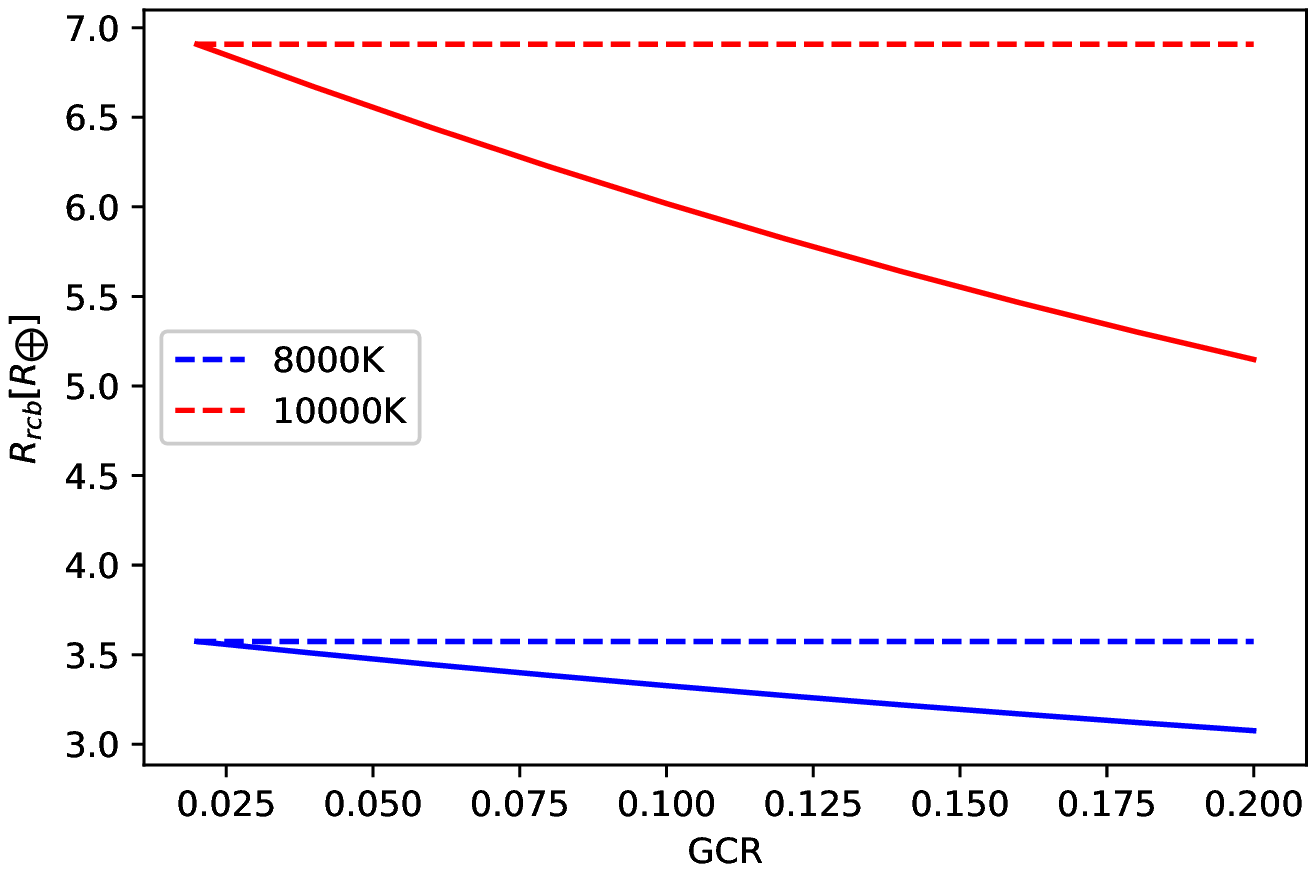}
  \includegraphics[width=7.42cm]{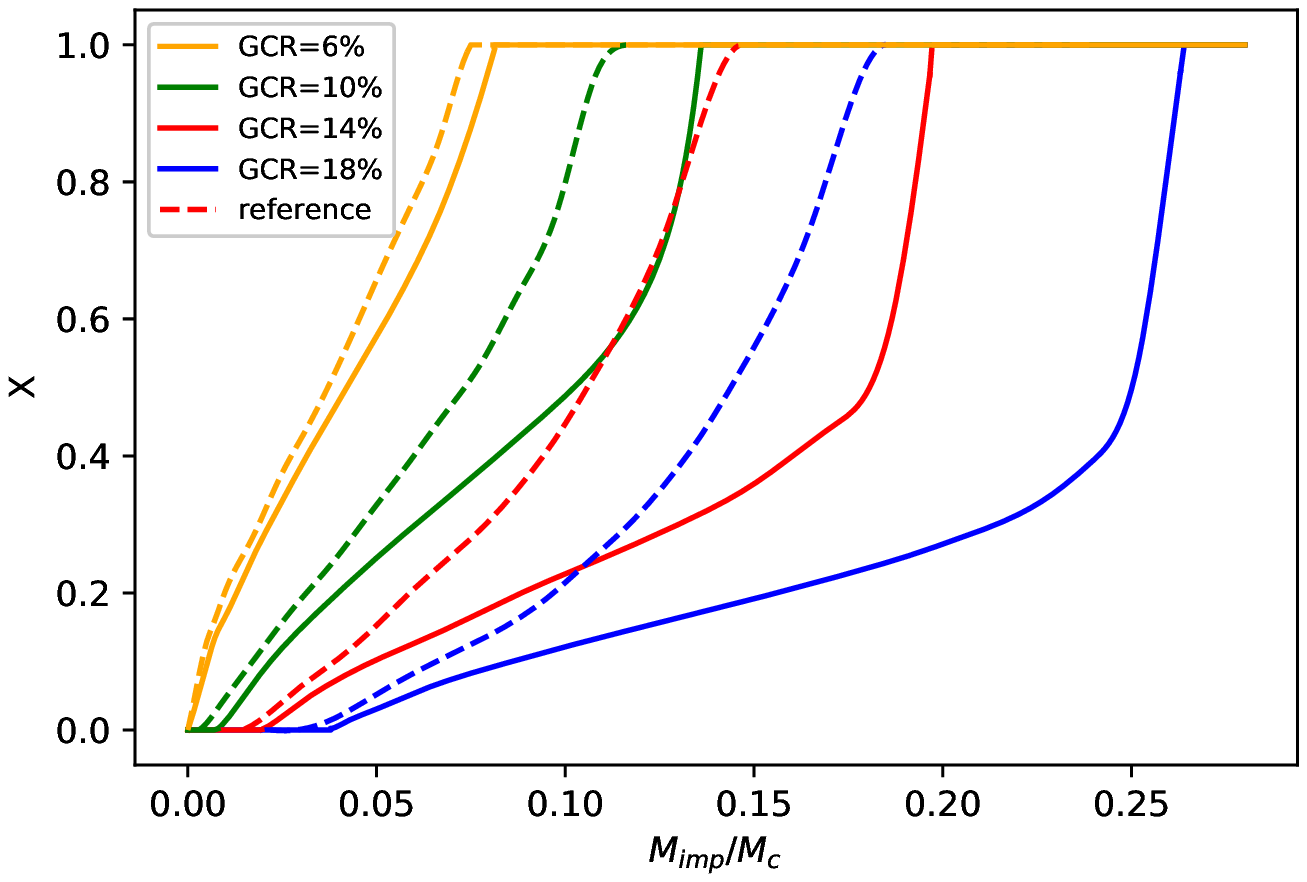}

   \caption{The left panel: the location of RCB ($R_{\rm rcb}$) is a function of gas-to-core ratio GCR. Blue and red lines correspond to initial base temperatures of $8000$ and $10,000 K$, respectively. The location of the RCB depends on the base temperature for the case without self gravity. The right panel: mass fraction of H/He envelope loss ($X$) is a function of impactor mass ($M_{imp}/M_c$). Different colors indicate an increase in the initial GCR from left to right. The solid and dashed lines represent the evolution with/without self-gravity, respectively. }
   \label{Fig:2}

   \end{figure}

Self-gravity is essential in the formation of planets. In section \ref{RCB}, we focus on the effects of self-gravity on the evolution of the position of the RCB.  The effects of self-gravity on the mass of the impact object are listed in section \ref{impactor}.

\subsubsection{The locations of the RCBs }\label{RCB}

If the self-gravity of the envelope is not included, the location of the RCB is completely determined by the base temperature $T_b$. Taking the self-gravity of the envelope into account, the RCB will move inward. The radius at RCB decreases with the increasing in the initial temperature at the base of the envelope or the initial GCR (as seen in the left panel of Figure \ref{Fig:2}).

\subsubsection{The mass of the impactor}\label{impactor}

The essence of giant impact is to inject energy into the planet. The planetary energies will be derived with a given impactor mass. The impactor mass and speed are represented by $M_{\rm imp}$  and $v_{\rm imp}$, respectively. The efficiency of energy conversion is described by $\eta$. The energy, induced by the impactor, is written as follows:

\begin{equation}\label{equation:20}
E_{\rm imp}=\eta \frac{M_{\rm imp} v_{\rm imp}^2}{2}.
\end{equation}
The efficiency of energy conversion $\eta$ approximates one as the kinetic energy for the impactor is completely transferred to the planet. As $v_{\rm imp}\approx v_{\rm esc}$ ($v_{\rm esc}$ is is the escape velocity of the planet), the above Equation will change to \citep{2019ESS.....430005B}

\begin{equation}\label{equation:21}
E_{\rm imp}=\eta \frac{GM_{\rm p} M_{\rm imp}}{R_{\rm c} }.
\end{equation}
The escape velocity changes with the atmospheric mass. In order to eliminate the effects of escape velocity, here, the atmospheric mass is neglected in our calculations, $M_{\rm p}\approx M_{\rm c}$. Equation (\ref{equation:21}) will switch into

\begin{equation}\label{equation:22}
E_{\rm imp}=\eta \frac{GM_{\rm c} M_{\rm imp}}{R_{\rm c} }.
\end{equation}
 When the impactor hits the planet with mass $M_{\rm imp}$, the energy transferred to the planet can be ensured. This energy will heat the planet and determine a new initial state of planet.

 \begin{figure}
   \centering
  \includegraphics[width=7.42cm]{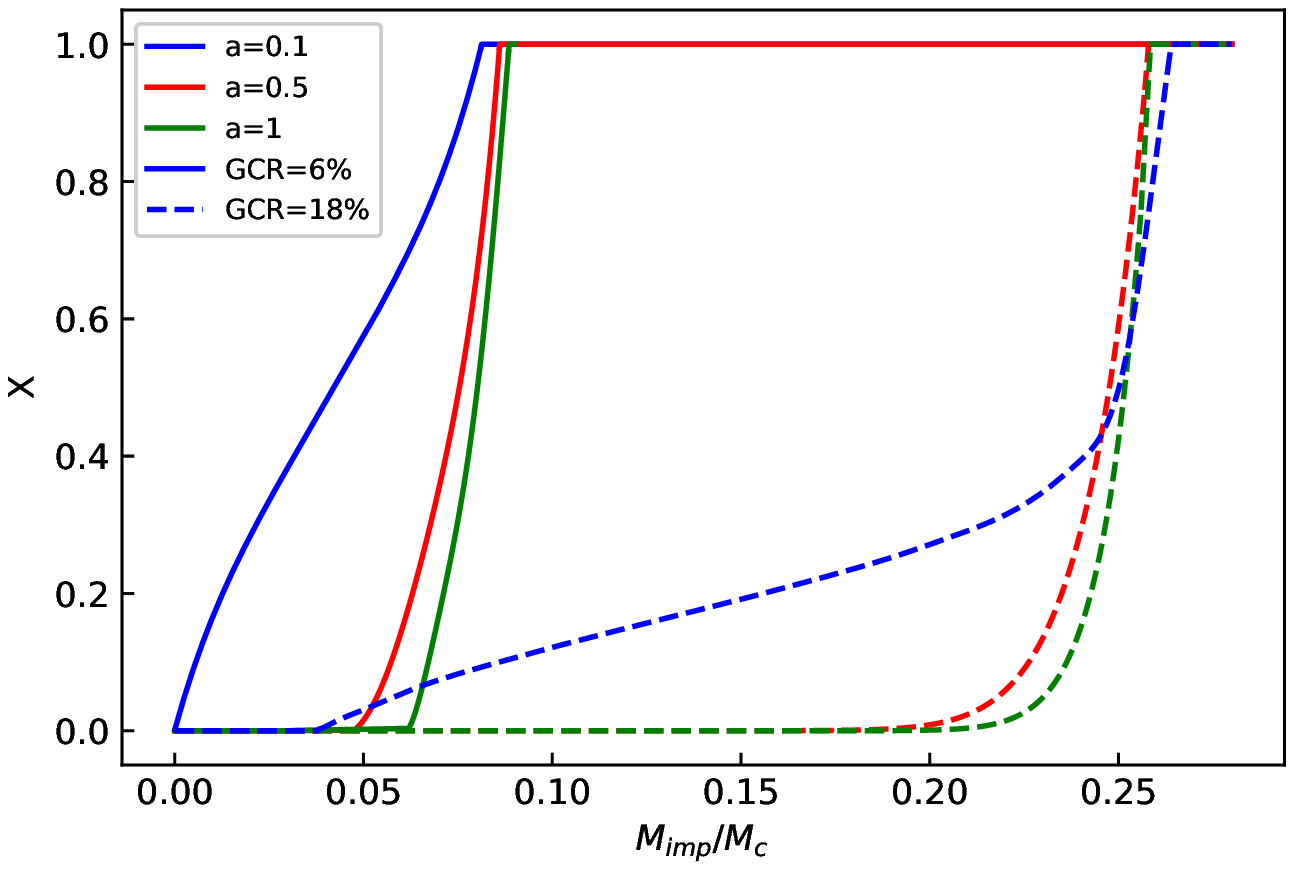}
  \includegraphics[width=7.42cm]{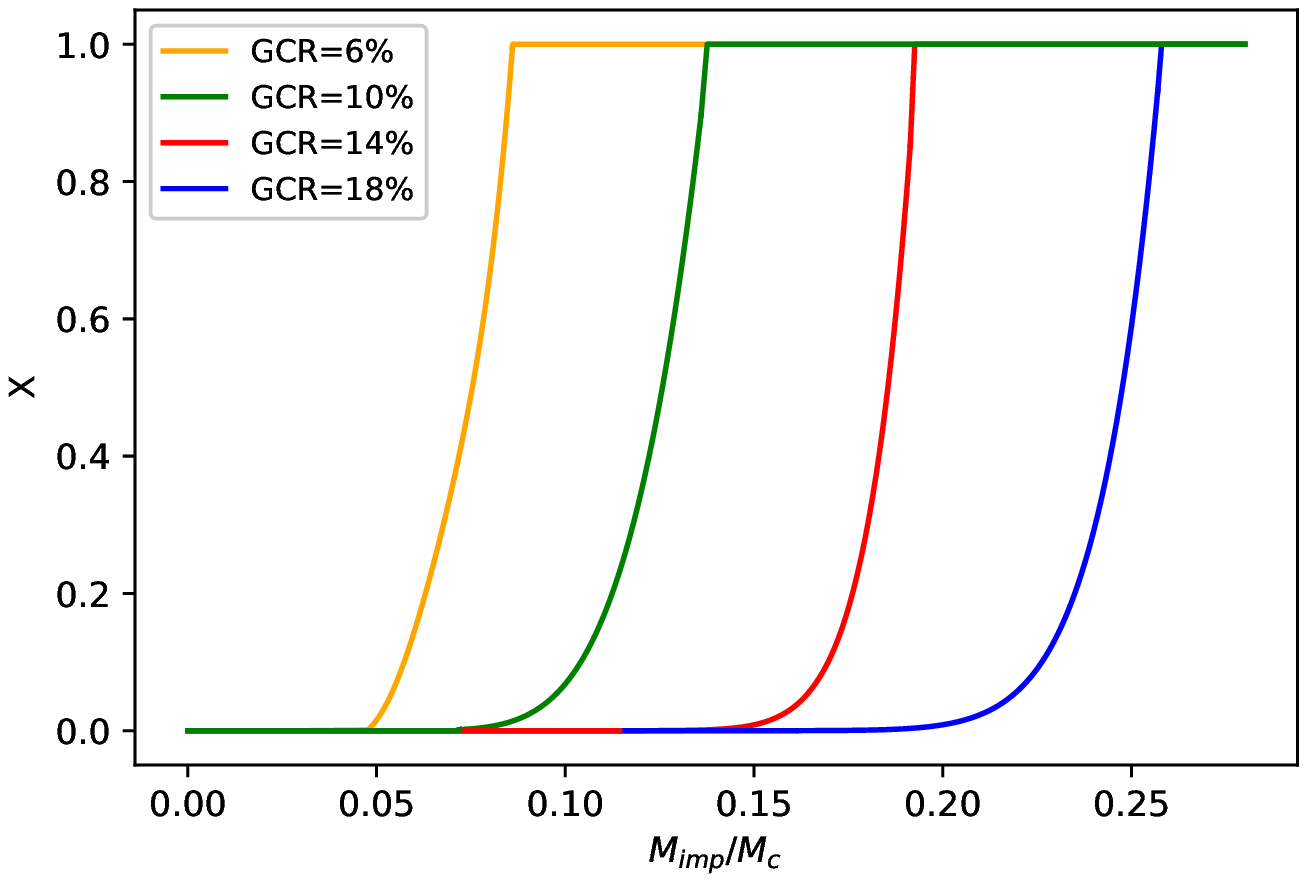}

   \caption{Mass fraction of H/He envelope loss ($X$) is a function of impactor mass ($M_{\rm imp}/M_{\rm c}$). The left panel is under the different orbital radius. Solid and dashed lines correspond to the cases for $GCR = 6\%$ and $18\%$. The right panel is under the different initial GCR and the planets are located at $0.5 au$.}
   \label{Fig:5}

  \end{figure}

The mass fraction of H/He envelope loss ($X$), which is the ratio of the retained envelope mass to the initial mass when the planetary evolution time is  $2 Gyr$, is a function of the impactor mass and shown in the right panel of Figure \ref{Fig:2}. The initial core temperature is $T_{\rm b,0}=T_{\rm c,0}= $4000K. The adiabatic index and the mean molecular weight of the planetary envelope satisfy $\gamma=7/5$, $\mu=2.3$ $u$. If the impactor mass approximates the envelope mass, the atmospheric mass for the planet without the self-gravity will be lost a lot \citep{2019ESS.....430005B}.   It is well known that self-gravity increases with the atmospheric mass and the core mass. In this section, the core mass is fixed.  When the initial atmospheric mass is small($GCR< 0.1$), the influence of self-gravity is weak.  When $GCR> 0.1$,  a larger impactor mass is required for the significant mass loss. As mentioned above, the RCB will be pushed inward by the self-gravity, implying the initial temperature required to produce significant mass loss is higher.

\subsection{The effect of the orbital radius}

In general, the close-in planets are more likely to form super-Earths because the host star affects them. We investigate the effects of orbital radius on the mass loss for the planets with self-gravity (seen in Figure \ref{Fig:5}). Following Equation (\ref{t_eq}), the equilibrium temperature will decrease with the orbital radius, resulting in the inwardly decrease in $R_{\rm RCB}$. The outer boundary will obtain stronger constrains by the greater self-gravity. A higher temperature is needed for the significant mass loss of the envelope with the larger impact mass. In addition, the whole envelope may be blown away. With the increase of $a$, atmospheric loss is quite sensitive to the change of impactor mass. There is a significant difference in mass-loss when the impactor mass changes slightly.

\section{DISCUSSION AND CONCLUSIONS}\label{sect:conclusions}

We construct a post-impact envelope model with an interior adiabatic convective zone and an exterior radiative zone. The thermal structure is governed by base temperature $T_{\rm b}$ and gas-to-core ratio GCR. The envelope self-gravity move the RCB inward, changing the envelope distribution. With the increase of $T_{\rm b}$ and GCR, the signature becomes more prominent. When $T_{\rm b}$ is relatively large, the atmosphere will have a significant mass-loss. Thus, self-gravity has a greater effect on large mass-loss, which helps the atmosphere retain more mass.

Giant impact is a process of recharging the planet. In the process, the kinetic energy of the impactor is converted into heat transferred to the planet. The core and envelope will be heated, determining a new planet state. This state works as the initial value of the evolutionary model constructed in Section \ref{sect:structure and evolution}, and different impactor mass corresponds to different initial values and evolutionary results. For the case with/without envelope self-gravity, evolutionary results are significantly different, especially in the case of large GCR. When the impactor mass and envelope mass are roughly equal, the envelope can experience a large amount of mass-loss, and even completely remove the envelope. After adding envelope self-gravity, gas is more constrained by the planet, which puts forward a stricter demand on impactor mass. Close-in planets are more prone to mass-loss. However, distant planets are more sensitive to changes in impactor mass. While it is more difficult to experience mass-loss, it is easy to remove the envelope completely once it does.

In addition to the atmospheric loss caused by thermal aspect, there is an atmospheric loss from impact-generated shocks \citep{2015Icar..247...81S}, which will have an effect on the envelope structure in the evolutionary process, but the mass loss caused by the thermal aspect plays a leading role, so in the paper we ignore it. We also ignore the photoevaporation \citep{2017ApJ...847...29O}, because, for planets with larger semimajor axes, photoevaporation is less important, and the effect is negligible compared with the mass-loss rate caused by giant impact.

\normalem
\begin{acknowledgements}

This work has been supported by the National Key R\&D Program of China (No. 2020YFC2201200) and the science research grants from the China Manned Space Project (No. CMS-CSST-2021-B09 and CMS-CSST-2021-A10). C.Y. has been supported by the National Natural Science Foundation
of China (grants 11373064, 11521303, 11733010, and 11873103), Yunnan National Science Foundation (grant Q9 2014HB048), and Yunnan Province (2017HC018).
\end{acknowledgements}

\bibliographystyle{raa}
\bibliography{bibtex}

\end{document}